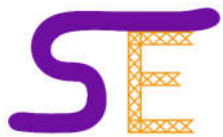

# Assessing the Potential of Interactive Vision Videos for Requirements Engineering

*Work Package WP2*
**Develop Interdisciplinary Guideline**

*Deliverable D1.0*
**An Interdisciplinary Guideline for the Production of Videos and Vision Videos by Software Professionals**

| | | |
|---|---|---|
| *Funding Instrument:* | Deutsche Forschungsgemeinschaft (DFG) | |
| *Project Duration:* | 2017 – 2020 | |
| *Prepared by:* | Oliver Karras and Kurt Schneider | |

| **Dissemination Level** | | |
|---|---|---|
| PU | Public | ✓ |
| PP | Restricted to other programme participants (including the Commission Services) | |
| RE | Restricted to a group specified by the Consortium (including the Commission Services) | |
| CO | Confidential, only for members of the Consortium (including the Commission Services) | |

## Deliverable Information

| Document Administrative Information | |
|---|---|
| Project Acronym: | ViViReq |
| Project Number: | 289386339 |
| Deliverable Number: | D1.0 |
| Deliverable Full Title: | An Interdisciplinary Guideline for the Production of Videos and Vision Videos by Software Professionals |
| Deliverable Short Title: | Guideline for (Vision) Video Production |
| Prepared by: | Oliver Karras and Kurt Schneider |
| Report Version: | v1.1 |
| Report Submission Date: | 15/03/2021 |
| Dissemination Level: | PU |
| Nature: | Technical Report |
| Lead Author(s): | Oliver Karras |
| Co-author(s): | Kurt Schneider |

## Change Log

| Date | Version | Author/Editor | Summary of Changes Made |
|---|---|---|---|
| 18/01/2020 | v1.0 | Oliver Karras | Initial version |
| 15/03/2021 | v1.1 | Oliver Karras | Minor revision |

# Summary

**Background and Motivation:** In recent years, the topic of applying videos in requirements engineering has been discussed and its contributions are of interesting potential. In the last 35 years, several researchers proposed approaches for applying videos in requirements engineering due to their communication richness and effectiveness. However, these approaches mainly use videos but omit the details about how to produce them. This lack of guidance is one crucial reason why videos are not an established documentation option for successful requirements communication and thus shared understanding. Software professionals are not directors and thus they do not necessarily know what constitutes a good video in general and for an existing approach. Therefore, this lack of knowledge and skills on how to produce and use videos for visual communication impedes the application of videos by software professionals in requirements engineering.

**How to Create Effective Videos and Vision Videos?:** This technical report addresses this lack of knowledge and skills by software professionals. We provide two guidelines that can be used as checklists to avoid frequent flaws in the production and use of videos respectively vision videos. Software professionals without special training should be able to follow these guidelines to achieve the basic capabilities to produce (vision) videos that are accepted by their stakeholders. These guidelines represent a core set of those capabilities in the preproduction, shooting, postproduction, and viewing of (vision) videos. We do not strive for perfection in any of these capabilities, .e.g., technical handling of video equipment, storytelling, or video editing. Instead, these guidelines support all steps of the (vision) video production and use process to a balanced way.

We have to remark that a vision video is a specific kind of video that represents a vision or parts of it for achieving shared understanding among all parties involved by disclosing, discussing, and aligning their mental models of the future system.

**Research Project ViViReq:** As part of the research project "Assessing the Potential of Interactive Vision Videos for Requirements Engineering (ViViReq)", we conducted a literature review of generic video production guidelines in terms of books and grey literature from several disciplines such as sociology, filmmaking, and video production. The recommendations of these generic video production guidelines have been discovered through years of experience. Thus, they represent a grounded and reflected body of knowledge on how to produce and use a good video. The data analysis of the found guidelines resulted in 307 text passages representing guiding recommendations for the production and use of generic videos. We analyzed, grouped, and merged these text passages resulting in a final set of 63 recommendations forming our interdisciplinary guideline for the production and use of generic videos by software professionals.

**Combining Lessons Learned from Literature and Our Own Experience:** Based on the interdisciplinary guideline for generic videos, we derived an experienced-based guideline for the production and use of vision videos for requirements engineering due to our project-specific context. Based on our theoretical knowledge and practical experience in the production and use of vision videos in requirements engineering, we adopted and revised several recommendations of the interdisciplinary guideline whose usefulness we were able to confirm for vision videos. Currently, we included 54 recommendations whose usefulness we could confirm for the production and use of vision videos. In addition, we also added five new recommendations that we identified ourselves during the production and use of vision videos. From these 54 recommendations, we derived 47 recommendations for vision videos by combining some of the recommendations for generic video into individual recommendations for vision videos. We also added five new recommendations that we identified ourselves during the production and use of vision videos. Thus, we present an experience-based guideline for vision video production that consists of 52 individual recommendations.

**Guidelines Ready to Use:** These two guidelines are a light-weight approach to increase the acceptance and value of (vision) videos by software professionals and their stakeholders. The guidelines are supposed to help software professionals in achieving the fundamental knowledge and skills to produce and use good (vision) videos at moderate costs and sufficient quality. Each recommendation advises the reader on how to proceed in the (vision) video production to achieve a particular result or outcome for a (vision) video. We also provide a rationale for each recommendation to explain why the particular recommendation is given and should be considered. Furthermore, we assigned one or more affected steps of the (vision) video production and use process as well as one or more affected quality characteristics of (vision) videos to each recommendation.

# Structure of the Technical Report

This document provides two guidelines that can be used as checklists to avoid frequent flaws in the production and use of videos respectively vision videos. Section 1 presents the motivation and background of this technical report. In Section 2, we explain the scientific deduction of the interdisciplinary guideline from literature with its 63 identified recommendations. Section 3 contains the experience-based guideline with its 52 recommendations based on the scientifically extracted recommendations from the literature and our own experience. Finally, we present each guideline as a comprehensive list in the Appendix A – Video Production Guideline and Appendix B – Vision Video Production Guideline.

# Table of Contents

# List of Figures

# List of Tables

# 1 Introduction

This section describes the target population of the two guidelines. We also provide a brief overview of the general video production and use process with its individual steps. Furthermore, we briefly explain (vision) video quality by introducing a quality model for (vision) videos and the high-level intended purposes of (vision) videos.

## 1.1 Target Population of the two Guidelines

The two guidelines for (vision) video production and use are intended to be used by software professionals who want to produce (vision) videos in the context of their software projects. Nevertheless, we assume that these set of recommendations can be useful for anyone who wants to produce a (vision) video. Especially, the recommendations of the interdisciplinary guideline represent the essence of the analyzed literature that forms a grounded and reflected body of knowledge on how to produce a good video.

## 1.2 Video Production and Use Process

Figure 1 shows the typical video production and use process that consists of four main steps called *preproduction*, *shooting*, *postproduction*, and *viewing* [3, 19]. Each of these steps should add value to a video. However, insufficient performance or violation of video quality in any of these steps can limit the final value of a video and thus the acceptance of the video by the audience. Insufficient quality in an early step will often constrain the quality that can be achieved in the later steps. Therefore, any early flaw decreases and limits the possible value of the final video.

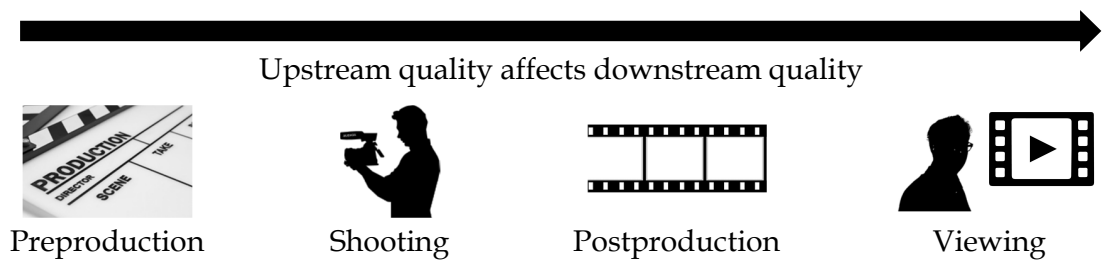

Figure 1: Video Production and Use Process; based on Karras et al. [16]

Below, we briefly explain the single steps of the video production and use process.

**1. Step: Preproduction**
In the preproduction, the planning of the video takes place by clarifying preliminaries, preparations, and the organization before the shooting begins. In this phase, the video producer has to define the purpose of a video, its story and single scenes, the desired duration, target audience, and the relevant contents that need to be shown. This preparation is necessary to ensure that the shooting can be done quickly and easily.

**2. Step: Shooting**
In the shooting, the video is recorded. The video producer has to record a shot for each scene in the story. These single shots are later combined into one video during the postproduction. In this phase, the video producer has to focus on capturing the relevant contents in pleasant way by avoiding disrupting and distracting contents, e.g., background noise or actions.

### 3. Step: Postproduction

In the postproduction, the whole video is created by combining and possibly rearranging the single shots to convey the entire planned story. Thereby, the image and sound of the composed video is edited and digitally post-processed to remove disrupting and distracting contents that were accidentally recorded. Besides improving the quality of the final video, the video producer has to ensure that the final video clearly presents the relevant contents in a pleasant and interesting for the target audience.

### 4. Step: Viewing

In the viewing, the entire video exists and is viewed as a whole by its target audience. The purpose of the video needs to be clearly communicated to the target audience to ensure that the viewers understand the value of the video. If the target audience does not understand or know the purpose of a video presented, it is possible that they reject the video since they do not recognize its value.

## 1.3 Video Quality

Video quality is a complex concept numerous factors [26]. On the one hand, there are technical factors, e.g., video properties (resolution, brightness, etc.) and record and display devices. On the other hand, there are subjective factors, e.g., individual interests, expectations, and experiences of viewers. The wide variety and subjectivity of these factors indicate the complexity of video quality. This complexity impedes the prediction of how different viewers assess the quality of a video due to several factors that affect their attitude [10]. Inspired by ISO/IEC FDIS 25010:2010 [6] for system and software quality models, Karras et al. [16] developed a quality model for videos (see Figure 2). This quality model can be used to identify relevant quality characteristics that can be further used to establish requirements, their criteria for satisfaction, and corresponding measures for a particular video. The quality model for videos by Karras et al. [16] divides the individual quality characteristics into the three dimensions of a quality model: representation, content, and impact. The representation dimension includes the sensorial characteristics of a video. The content dimensions includes the perceptual characteristics of the content of a video. The impact dimensions includes the emotional characteristics regarding the impact of a video on its target audience. Below, the single quality characteristics of videos are briefly explained. For further information, we refer to Karras et al. [16].

**Dimension: Representation.** This dimension covers the sensorial characteristics of a video regarding its representation (see Table 1). The characteristic *video stimuli* aggregates the three sub-characteristics *image quality*, *sound quality*, and *video length* which a viewer perceives with his sensory organs.

*Table 1: Dimension: Representation – Video stimuli*

| |
|---|
| **Dimension:** <br>    *Representation* covers the sensorial characteristics of a video. |
| **Characteristic:** <br>    *Video stimuli* considers the sensorial stimuli of a video. <br> **Sub-characteristics:** <br>    *Image quality* considers the visual quality of the image of a video. <br>    *Sound quality* considers the auditory quality of the sound of a video. <br>    *Video length* considers the duration of a video. |

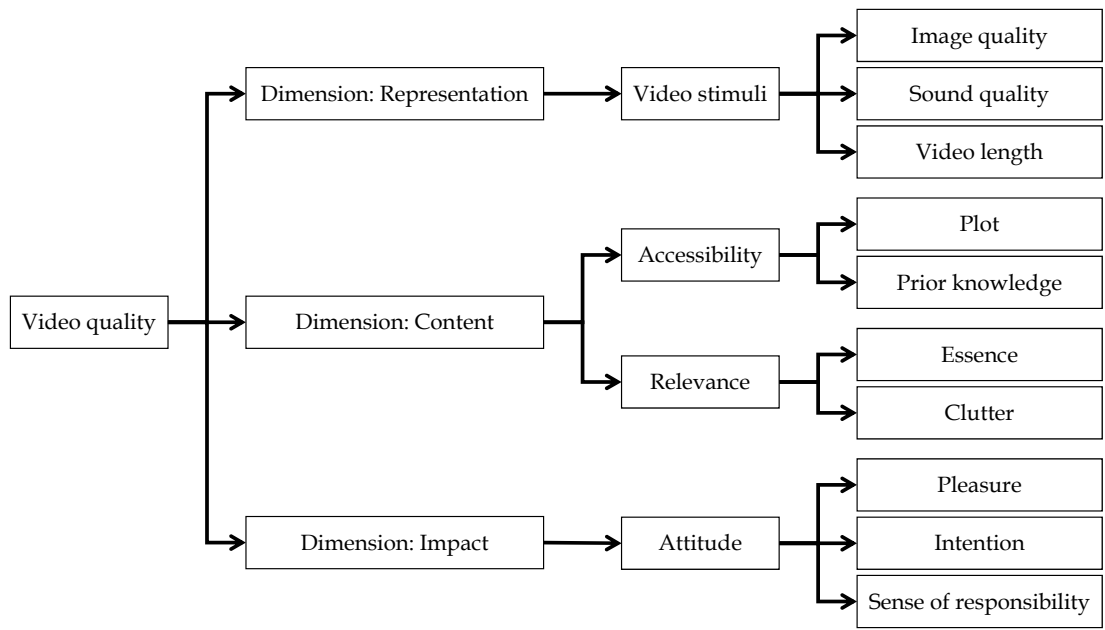

*Figure 2: Hierarchical Decomposition of Video Quality*

**Dimension: Content.** This dimension covers the perceptual characteristics of a video regarding its content (see Table 2). The content dimension includes the two characteristics *accessibility* and *relevance* which are further subdivided. *Accessibility* focuses on the ease of access to the content of video by containing the sub-characteristics *plot* and *prior knowledge*. A video is easier or harder to access depending on the structuring of the content and the presupposed prior knowledge. *Relevance* considers the presentation of valuable information in a video by including the sub-characteristics *essence* and *clutter*. These two sub-characteristics distinguish between important core elements, e.g., persons and locations, as well as disrupting and distracting elements, e.g., background actions and noises.

*Table 2: Dimension: Content – Accessibility and relevance*

| |
|---|
| **Dimension:** <br>   *Content* covers the perceptual characteristics of a video. |
| **Characteristic:** <br>   *Accessibility* considers the ease of access to the content of a video. <br> **Sub-characteristics:** <br>   *Plot* considers the structured presentation of the content of a video. <br>   *Prior knowledge* considers the presupposed prior knowledge to understand the content of a video. |
| **Characteristic:** <br>   *Relevance* considers the presentation of valuable content in a video. <br> **Sub-characteristics:** <br>   *Essence* considers the important core elements, e.g., persons, locations, and entities, which are to be presented in a video. <br>   *Clutter* considers the disrupting and distracting elements, e.g., background actions and noises, that can be inadvertently recorded in a video. |

**Dimension: Impact.** This dimension covers the emotional characteristics of a video regarding its emotional impact on its target audience (see Table 3). The characteristic *attitude* includes the sub-characteristics *pleasure*, *sense of responsibility*, and *intention*. *Pleasure* focuses on the enjoyment of watching a video. *Sense of responsibility* deals with the fulfilment of legal regulations. *Intention* considers the intended purpose of a video which defines the use of a video and why a video is necessary. The high-level intended purposes of a video are described in the subsequent Section 1.4.

Table 3: Dimension: Impact – Attitude

| |
| --- |
| **Dimension:**<br>    *Impact* covers the emotional characteristics of a video. |
| **Characteristic:**<br>    *Attitude* considers the humans' conception of a video.<br>**Sub-characteristics:**<br>    *Pleasure* considers the enjoyment of watching a video.<br>    *Intention* considers the intended purpose of a video.<br>    *Sense of responsibility* considers the compliance of a video with the legal regulations. |

## 1.4 Intentions of Videos

The intended purpose of a video explains the reason and immediate goal behind the use of a video and thus represents the underlying intention of a video. It is important to know why a video is necessary to satisfy the information needs of the target audience. According to Hanjalic et al. [5], there are the following five high-level intended purposes of videos.

**1. Information:** Convey or obtain knowledge and/ or new information (declarative knowledge). This intended purpose covers the cases which have the goal of conveying or obtaining declarative knowledge, i.e., 'knowing that'.

**2. Experience learning:** Convey or obtain skills or something practically by experience (procedural knowledge). This intended purpose covers knowledge acquisition regarding skills or procedural knowledge, i.e., 'knowing how'.

**3. Experience exposure:** Convey or obtain particular experiences. The video serves as a replacement of an actual person, place, entity, or event. This intended purpose covers the cases which have the goal of conveying or obtaining exposure to real-life or another type of experience.

**4. Affect:** Convey or obtain a mood or affective state. The video serves for relaxation or entertainment purposes. This intended purpose covers the cases which have the goal of conveying or obtaining a mood, affective, or physical state including relaxation and in general other effects of entertainment.

**5. Object:** Convey or obtain content in form of a video to serve a particular purpose in a real-life situation. This intended purpose covers the cases which have the goal of conveying or obtaining a video suitable for a particular real-world situation.

The first four intended purposes have two generalizations in common. First, there is probably a large convergence between the producer's and viewer's main intended purpose of a video, e.g., a tutorial is made mainly to teach people skills. Second, the goals of these intended purposes can also be accomplished without a video, e.g., skills can be acquired by taking lessons from a teacher. The difference between these four intended purposes becomes more obvious by asking which real-world activity can help to achieve the same goal.

The fifth intended purposes differs from the other ones in both generalizations. First, the producer's and viewer's intended purpose might diverge radically from each other. This divergence is caused by the difference in the second generalization. The fifth intended purpose covers the goals that necessarily require a video and thus cannot be accomplished by other means. Therefore, the viewer looks for a video that fulfills his needs and intention which do not have to match with the producer's ones. This mismatch is possible since the five high-level intended purposes are not mutually exclusive but overlap [5].

## 1.5 Vision Video Quality

The following sections introduce vision video quality which needs to be explained to facilitate the understanding of the experience-based guideline for vision videos. Karras et al. [16] also developed a quality model for visions (see Figure 3) which they combined with their quality model for videos to derive a quality model for vision videos (see Figure 13).

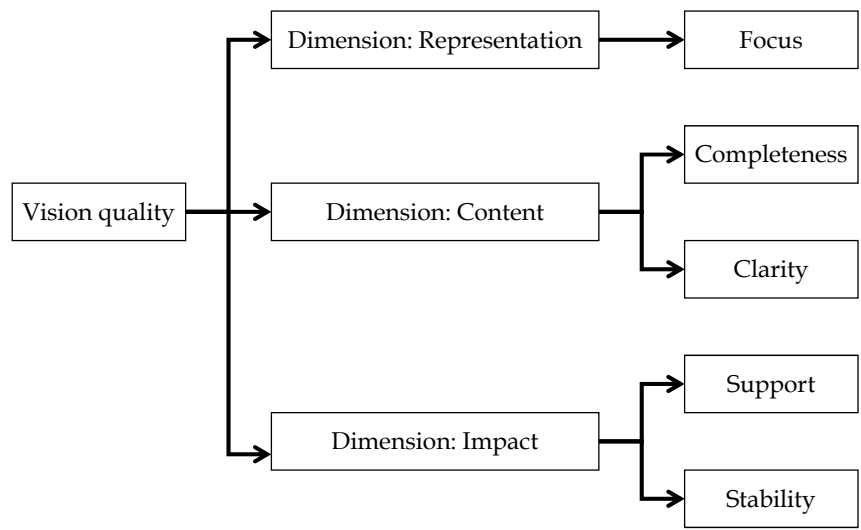

Figure 3: Hierarchical Decomposition of Vision Quality

### 1.5.1 Vision Quality

Below, we briefly explain the quality characteristics of visions. For further information, we refer to Karras et al. [16].

**Dimension: Representation.** This dimension covers the sensorial characteristics of a vision regarding its representation (see Table 4). The characteristics *focus* considers the condensed and short description of a vision by presenting its essence compactly. The description of a vision can be either a text (a few sentences up to one page) or a picture. The vision must shows the conceptual image of the future product as a "big picture".

Table 4: Dimension: Representation – Focus

| | |
|---|---|
| **Dimension:** | |
| *Representation* covers the sensorial characteristics of a vision. | |
| **Characteristic:** | |
| *Focus* considers the compact representation of a vision. | |

**Dimension: Content.**  This dimension covers the perceptual characteristics of a vision regarding its content (see Table 5). The content dimension includes the two characteristics *completeness* and *clarity*. A vision consists of three contents: The addressed *problem*, the key idea of the *solution*, and how the solution *improves* the state-of-the-art. *Completeness* deals with the coverage of these three contents of a vision. An understandable vision depends on the ability of a company to define clear objectives. Otherwise, the product development delays since ambiguous concepts of a product allow speculations and conflicts about should be produced. *Clarity* considers the intelligibility of the aspired goals of a vision by all parties involved.

Table 5: Dimension: Content – Completeness and clarity

| | |
|---|---|
| **Dimension:** | |
| *Content* covers the perceptual characteristics of a vision. | |
| **Characteristic:** | |
| *Completeness* considers the coverage of the three contents of a vision, i.e., *problem*, *solution*, and *improvement*. | |
| *Clarity* considers the intelligibility of the aspired goals of a vision by all parties involved. | |

**Dimension: Impact.**  This dimension covers the emotional characteristics of a vision regarding its emotional impact on all parties involved (see Table 6). The impact dimension includes the two characteristics *support* and *stability*. A vision needs support in the development team and has to reflect a balanced view that satisfies the needs of diverse stakeholders. *Support* focuses on the level of acceptance of vision, i.e., whether all parties involved share and accept the same vision as their motivation and guidance of their actions and activities. A vision needs to be stable with consistent objectives over time to provide consistency to short-term actions while leaving room for reinterpretation as new opportunities emerge. Therefore, *stability* focuses on the ability of an organization to learn and adapt to finish a project successfully.

Table 6: Dimension: Impact – Support and stability

| | |
|---|---|
| **Dimension:** | |
| *Impact* covers the emotional characteristics of a vision. | |
| **Characteristic:** | |
| *Support* considers the level of acceptance of a vision, i.e., whether all parties involved share and accept the vision. | |
| *Stability* considers the consistency of a vision over time. | |

### 1.5.2  Intentions of Visions

As for videos, a vision also has its intended purpose that explains the reason and immediate goal behind its use. According to Karras et al. [16], there are the following three intended purposes of visions.

1. *Share an integrated view of a future system and its use within a heterogeneous group of stakeholders to align their actions and views.*

A vision synthesizes the contributions, findings, and implications solicited from stakeholders. The vision serves as a guiding principle for all stakeholders to align their actions.

2. *Share an integrated view of a future system and its use with the development team that will implement the vision.*

A vision is a "guiding star" for the project members to develop a productive business response for the stakeholders since it provides the context for decision-making. This support is especially important in the initial phase of a project when a team conceives and develops the first version of a product.

3. *Share an integrated view of a future system and its use for validating this view and for eliciting new or diverging aspects.*

The requirements engineering process needs to establish a vision in the relevant system context. Therefore, an integrated view of the potential future system and its use is necessary to validate whether different stakeholders share this view. In the case of inconsistencies or ambiguities, all parties involved can negotiate to make joint decisions to conceive and develop a satisfying system.

The first two intended purposes of visions are related to the intended purpose of videos: Information. In both cases, declarative knowledge is shared, whereby only the respective target audience is different. The third intention corresponds to the intended purpose of videos: Experience Exposure. In this case, the video serves as a replacement of the future system and its use so that the viewers can experience the envisioned product.

### 1.5.3 A Quality Model for Vision Videos

Karras et al. [16] obtained their final quality model for vision videos by combining the two individual quality models for vision and video. Figure 4 presents the quality model for vision videos whose labels follow the previous explanations and definitions.

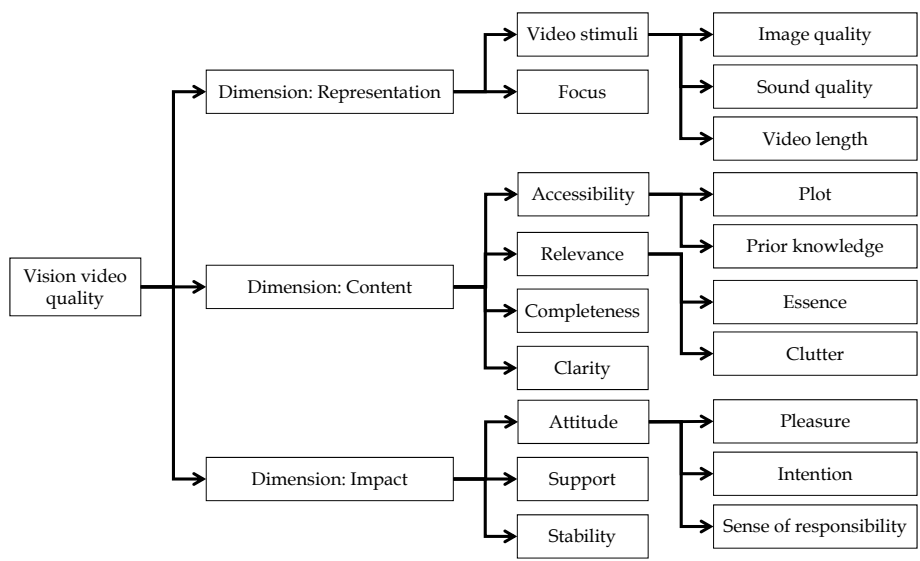

*Figure 4: Hierarchical Decomposition of Vision Video Quality*

# 2 Detailed Video Production Guideline

In this section, we present the 63 recommendations of our interdisciplinary guideline for video production and use by software professionals. In the following, we present each individual recommendation in detail, i.e., the recommendation, its rationale, the affected steps of the video production and use process as well as the affected quality characteristics of videos. In Appendix A – Video Production Guideline, we present the interdisciplinary video production guideline as a comprehensive list that only contains the 63 recommendations.

## 2.1 Development of the Interdisciplinary Guideline

Based on 307 extracted text passages of generic video production guidelines (cf. Karras et al. [16]), we developed the interdisciplinary video production guideline. Three computer science researchers with an average more than three years of experience in video production in requirements engineering assessed each text passage in a two-hour workshop based on their experience. In the workshop, the three researchers discussed each text passage whether the text passage represents some kind of recommendation that suggests how something in the video production and use should be done (HOW) to achieve a particular result or outcome for a video (WHAT) by emphasizing the reason for considering this advice (WHY). These three aspects (HOW, WHAT, and WHY) are important for any kind of recommendation to understand why we do what we do [25]. The HOW defines the process or method used, the WHAT defines the result or outcome of the process or method, and the WHY defines the motivation, reason, or purpose for applying the process or method. Only when all three researchers agreed on the presence of all three aspects (HOW, WHAT, and WHY) the text passage was selected as a candidate for the interdisciplinary video production guideline.

As a result of the workshop, we identified 105 text passages as candidates which we grouped and merged resulting in a final set of 63 candidates. From these 63 text passages, we extracted the HOW, WHAT, and WHY manually to reformulate each candidate according to the following pattern: "[HOW] to [WHAT]. [WHY]." This pattern summarizes how someone should proceed (HOW) to achieve a particular result or outcome (WHAT) by providing the motivation, reason, or purpose for following the recommendation (WHY). Thus, we concretized each candidate as a structured recommendation resulting in a final set of 63 recommendations forming the interdisciplinary guideline for video production and use by software professionals.

We evaluated the content validity of all recommendations and the entire guideline by using the content validity index according to Polit et al. [21]. This index is a measure of the agreement among a number of raters who assess the individual items of an artefact whether these items and thus the artefact really provide the content that they are supposed to provide.

In particular, a group of five experts in video production in requirements engineering assessed each recommendation whether the recommendation provides the essential information about the process or method (HOW), the result or outcome (WHAT), and the motivation, reason, or purpose (WHY). Furthermore, the five experts assessed each recommendation regarding its relevance for inexperienced persons who want to produce a video. Based on the assessments of the individual recommendations, the same aspects can also be assessed for the entire interdisciplinary guideline to evaluate its respective content validity.

In summary, the average content validity index of all four aspects assessed yield an almost perfect raters' agreement that the recommendations and thus the interdisciplinary guideline are relevant for inexperienced persons, who want to produce a video, by really providing the essential information about the process or method, result or outcome, and motivation, reason or purpose. Thus, we are confident that we constructed an interdisciplinary guideline of fundamental relevance and soundness.

The following recommendations are sorted based on the affected steps of the video production and use process as far as possible since several recommendations affect more than one step. Thus, the recommendations are sub-sorted according to natural sequence of the video production and use process steps.

## 2.2 Detailed Overview of the Interdisciplinary Guideline

Below, we present each of 63 recommendation in detail, i.e., the recommendation, its rationale, the affected steps of the video production and use process as well as the affected quality characteristics of videos.

01. Have preliminary meetings with your stakeholders to become sensitive to their concerns and different viewpoints.
    **Rationale:** These meetings help you to understand better your stakeholders by establishing trust between all parties involved.
    **Steps:** Preproduction
    **Characteristics:** Sense of responsibility

02. Define the intended purpose and target audience of your video to plan the content of your video.
    **Rationale:** You must know why you want to produce a video and for whom to better plan what content you must include in your video.
    **Steps:** Preproduction
    **Characteristics:** Intention, Prior knowledge

03. Create a list of topics that you want to address in your video to define the final number of topics addressed in your video.
    **Rationale:** If you have too many topics, the audience rarely remembers more than a fraction of them. If you have too few topics, the audience perceives the video as slow and labored.
    **Steps:** Preproduction, Viewing
    **Characteristics:** Essence, Pleasure

04. Keep your video simple by addressing only a few topics to reduce the amount of information in your video.
    **Rationale:** The audience can understand a video with fewer topics easier than a video with too many topics.
    **Steps:** Preproduction, Viewing
    **Characteristics:** Essence, Pleasure

05. Deal with one topic at a time to avoid cuts between different topics, flashbacks, and flashforwards.
    **Rationale:** Any cut, flashback, or flashforward can confuse the audience making it difficult to understand your video.
    **Steps:** Preproduction, Postproduction, Viewing
    **Characteristics:** Essence, Plot, Pleasure

06. Ensure that the image- and soundtrack deal with the same topic to present consistent visual and auditive information.

| | **Rationale:** | The audience can be easily distracted and confused if the visual and auditive information of the image- and soundtrack do not match. |
| --- | --- | --- |
| | **Steps:** | Preproduction, Shooting, Postproduction, Viewing |
| | **Characteristics:** | Essence, Clutter, Pleasure |

**07.** Tell the content of your video as a story that has a beginning, middle, and end to create a clear structure of the contents of your video.
    **Rationale:** This structure helps you to ensure that you introduce the prior knowledge (beginning) which the audience needs to know to understand the content (middle) which has a defined conclusion (end).
    **Steps:** Preproduction, Viewing
    **Characteristics:** Plot, Prior knowledge

**08.** Use a storyboard, script, or narration to create an outline of the story of your video.
    **Rationale:** It is essential to plan the entire plot and main action of each scene in advance to organize the entire video production.
    **Steps:** Preproduction
    **Characteristics:** Plot, Essence

**09.** Plan the plot of your video by designing the scenes in such a way that they can be trimmed or omitted as needed to shorten the final video if necessary.
    **Rationale:** It may be necessary to reduce the duration of the final video afterward.
    **Steps:** Preproduction, Postproduction
    **Characteristics:** Video length, Essence, Plot

**10.** Think about how long a shot should last to define the final duration of the shot.
    **Rationale:** The duration of a shot is crucial. If a shot is too long, the audience loses interest since they cannot capture, process, and understand the information. If a shot is too brief, the audience captures, but cannot process and understand, the information.
    **Steps:** Preproduction, Shooting, Postproduction, Viewing
    **Characteristics:** Video length, Pleasure

**11.** Hold a shot for at least 15 seconds to enable the audience to understand the information presented.
    **Rationale:** Fifteen seconds is the lower boundary of the average storage period of the human short-term memory for capturing, processing, and understanding information.
    **Steps:** Preproduction, Shooting, Postproduction, Viewing
    **Characteristics:** Video length, Pleasure

**12.** Hold a shot for a maximum of 30 seconds to avoid too long shots that the audience cannot understand.
    **Rationale:** Thirty seconds is the upper boundary of the average storage period of the human short-term memory for capturing, processing, and understanding information. If you exceed this duration, it is difficult for the audience to capture, process, understand, and remember the important details of a shot.
    **Steps:** Preproduction, Shooting, Postproduction, Viewing
    **Characteristics:** Video length, Pleasure

**13.** Plan to shoot shots of strategic moments of a long action to show the important moments of the long action in a condensed shot.

> **Rationale:** You must keep the duration of a shot short but at the same time show all important moments of a long action regardless of its length.
> **Steps:** Preproduction, Shooting, Viewing
> **Characteristics:** Plot, Video length

14. Compose a shot by keeping the important details of a scene within the safe area, which is the 70 percent area around the center of the screen, to ensure that the subject is properly framed and you do not accidentally cut off important details of the scene.
    > **Rationale:** The audience might miss or not recognize the details that are outside of the safe area.
    > **Steps:** Preproduction, Shooting, Viewing
    > **Characteristics:** Image quality, Essence

15. Compose a shot by following the subsequent standards of the rule of thirds, which divides the screen into thirds horizontally and vertically, to create a shot that looks dynamic.

    The rule of thirds:
    1. The subject should not be exactly in the middle of the screen.
    2. The subject should be on one of those lines and, ideally, on the intersection of two lines.
    3. If the subject moves towards something, position the subject behind the center of the screen according to the direction of the camera motion.
    4. If the subject moves away from something, position the subject over the center of the screen according to the direction of the camera motion.
    5. The faster the movement of the subject, the greater the offset from the center of the screen.

    > **Rationale:** The rule of thirds helps you produce a nicely balanced image that attracts the interest of the audience.
    > **Steps:** Preproduction, Shooting, Viewing
    > **Characteristics:** Image quality, Essence, Pleasure

16. Compose a shot by including a suitable back- and foreground for a scene to add additional information and a meaningful context to your video.
    > **Rationale:** The back- and foreground must support the audience to understand the scene.
    > **Steps:** Preproduction, Shooting
    > **Characteristics:** Prior knowledge, Image quality, Essence

17. You have the subsequent options to obtain a suitable background for your scene.

    Options for the background of a scene:
    1. Use the real location.
    2. Use a substitute for the location you need.
    3. Build a set that resembles the real location.
    4. Combine photos from the real location and sound effects with your shots to make the audience think that you were shooting in the real location.

    > **Rationale:** It can be difficult or impossible to shoot in the real location. In this case, you need to think about alternatives to illustrate the scene of your video properly.
    > **Steps:** Preproduction, Shooting, Postproduction
    > **Characteristics:** Prior knowledge, Image quality, Essence

18. Ask the owner or responsible authority of private or public property to sign a consent form for shooting on the property to obtain the permission (signed consent form) to shoot on the property.
    **Rationale:** You are responsible and legally liable to comply with the legal regulations in the context of a video production.
    **Steps:** Preproduction, Shooting
    **Characteristics:** Sense of responsibility

19. Ask each actor to sign a consent form covering the subsequent topics to obtain the permission (signed consent form of each actor) to use and distribute the individual shots and the final video.

    Topics the consent form must cover:
    1. A statement of how all shots and the final video will be used and distributed.
    2. A statement that the actor has the right to withdraw at any time from the video production.
    3. A statement of whether the actor may request the deletion of the recordings at any time.
    4. A statement of how the actor can contact you (phone number or email).

    **Rationale:** You are responsible and legally liable to comply with the legal regulation of the general data protection regulation (GDPR).
    **Steps:** Preproduction, Shooting, Postproduction, Viewing
    **Characteristics:** Sense of responsibility

20. Give each actor a suitable amount of time for reading the consent form as well as asking questions and clarifying issues and concerns to achieve that the actor is completely informed about the video production and use.
    **Rationale:** You increase each actor's trust in you and your work by being transparent regarding the entire video production.
    **Steps:** Preproduction
    **Characteristics:** Sense of responsibility

21. Plan to shoot all of the action at one location before you go to the next location to simplify the shooting.
    **Rationale:** Thus, ou save time by not following the running order of the scenes according to the plot.
    **Steps:** Preproduction, Shooting
    **Characteristics:** Plot, Sense of responsibility

22. Follow the subsequent steps to record a shot.

    Steps for recording:
    1. Review the details of the storyboard for the next scene.
    2. Call "Quiet, please!".
    3. Start recording.
    4. Call "Action!" to start the action.
    5. Call "Cut!" to announce the end of the action.
    6. Stop recording.
    7. Review the recording. If you are unsatisfied with the shot, repeat the recording.

    **Rationale:** These steps are an established process in video production for making a recording.
    **Steps:** Shooting
    **Characteristics:** Essence, Clutter, Video length, Image quality, Sound quality

23. Handle inexperienced actors by following the subsequent rules of conduct to calm the actors down.

    Rules of conduct for handling inexperienced actors:
    1. Making the actors feel welcome and that their contribution is important for the video.
    2. Give them instructions by telling them when, where, and how they should act in front of the camera.
    3. Make it clear to the actors that they do not have to worry that something goes wrong since the scene can be shot again.

    **Rationale:** Inexperienced actors are often nervous and afraid of doing something wrong that impedes shooting.
    **Steps:** Shooting
    **Characteristics:** Sense of responsibility

24. Use the best camera available (smartphone, tablet, consumer HD camera, etc.) which is small and light and has a display, mounting option for an external microphone, a large storage capacity, and sockets for a connection with a computer to obtain high image and sound quality.
    **Rationale:** Today's video cameras offer an image and sound quality that is sufficient for the most purposes.
    **Steps:** Shooting
    **Characteristics:** Image quality, Sound quality

25. Use an external microphone, i.e., a shotgun microphone, to achieve high sound quality.
    **Rationale:** The built-in microphones rarely provide sufficient sound quality. A shotgun microphone is ideal for isolating a subject and eliminating nearby noises.
    **Steps:** Shooting
    **Characteristics:** Sound quality, Essence, Clutter

26. In general, let the camera control auto-focus, white balance, and exposure controls to obtain a high image quality.
    **Rationale:** These camera settings are complex and have a strong impact on the image. Therefore, you should only adjust them manually if you have the necessary knowledge and experience.
    **Steps:** Shooting
    **Characteristics:** Image quality

27. Use the 720p (1280 x 720) or better 1080p (1920 x 1080) HD format which utilizes a 16:9 aspect ratio to obtain a high-quality image.
    **Rationale:** These HD formats give your video a high-quality video look.
    **Steps:** Shooting
    **Characteristics:** Image quality

28. Use the standard recording speed to obtain high-quality image and sound.
    **Rationale:** The standard recording speed provides better image and sound quality.
    **Steps:** Shooting
    **Characteristics:** Image quality, Sound quality

29. Consider where you place the camera and microphone to create a video that enables the audience to experience the content of your video.
    **Rationale:** The more realistic the audience perceives the content of your video, the better they can understand the content of your video.

       **Steps:** Shooting, Viewing
       **Characteristics:** Image quality, Sound quality, Essence

30. When using the camera, keep in mind that you restrict the perspective of the image the audience can see in the video to avoid frustrating the audience.
    **Rationale:** The audience might get the impression that they are missing something.
    **Steps:** Shooting, Viewing
    **Characteristics:** Image quality, Essence, Pleasure

31. Have the light source, e.g., lamp or window, behind the camera to illuminate the subject.
    **Rationale:** The effect of light depends on the position of the camera. If you shoot against the light source, you leave the subject in deep shadow, making it difficult for the audience to recognize the subject.
    **Steps:** Shooting, Viewing
    **Characteristics:** Image quality, Essence

32. Place the microphone as close as possible to the subject by keeping the distance the same for all shots to get a higher sound quality.
    **Rationale:** The microphone is often too far away from the subject and strong reflections from nearby walls or loud background noises lower the quality of the sound.
    **Steps:** Shooting
    **Characteristics:** Sound quality, Clutter

33. For brief video production, shoot every action from start to finish to ensure that the audience misses none of the action.
    **Rationale:** The audience gets an accurate impression of the action of a scene and its duration.
    **Steps:** Shooting, Viewing
    **Characteristics:** Plot, Essence, Video length, Pleasure

34. Start the recording 5 – 10 seconds before the action starts and stop the recording 5 – 10 seconds after the action is completed to avoid too brief shots.
    **Rationale:** These additional buffers help you in the postproduction when you cut and combine the individual shots into the final video.
    **Steps:** Shooting, Postproduction
    **Characteristics:** Video length

35. Watch the elapsed time on the camera to know how much memory and battery are left.
    **Rationale:** A full memory and an empty battery delay the shooting, thus increase your costs, and waste the time of all parties involved.
    **Steps:** Shooting
    **Characteristics:** Sense of responsibility

36. Stabilize the camera by using your body or a camera mount (monopod or tripod) to create a steady and carefully controlled image of your video.
    **Rationale:** If the image of your video is blurred, bounce around, or lean over to one side, it is a pain to watch for the audience.
    **Steps:** Shooting, Viewing
    **Characteristics:** Image quality, Pleasure

37. When using a tripod, turn the auto-focus off. Instead, position the camera and manually focus the image on a fixed point of interest.

    **Rationale:** If you use the auto-focus, any movement in the scene may cause the camera to adjust and re-adjust which in turn blurs the image for a moment.
    **Steps:** Shooting
    **Characteristics:** Image quality

38. Use only as much camera motions, pans, zooms, and tilts as necessary by always ensuring that the motions are slow and smooth to create a shot with a low compression.
    **Rationale:** Any camera motion results in more compression of the image of a video and can cause motion sickness by the audience.
    **Steps:** Shooting, Viewing
    **Characteristics:** Image quality, Pleasure

39. If possible, do not zoom in or out to obtain a high image quality of your video.
    **Rationale:** Zooming is an unnatural eye movement that can confuse the audience.
    **Steps:** Shooting, Viewing
    **Characteristics:** Image quality, Pleasure

40. Sharpen the focus on the most important part of the scene and leave the rest defocused to highlight the important content of a scene.
    **Rationale:** Although this presentation style is not ideal, it enables you to show the audience exactly what you want them to see.
    **Steps:** Shooting, Viewing
    **Characteristics:** Image quality, Essence

41. Light the subject well by switching on the room lights or open the curtain to obtain a high image quality.
    **Rationale:** There is usually not enough light in a building to present a subject in such a way that the audience can recognize all details of the subject.
    **Steps:** Shooting, Viewing
    **Characteristics:** Image quality, Essence

42. Use a close shot to show the audience the action of a scene.
    **Rationale:** A close shot enables you to show the details of the scene by showing the audience the concrete action more closely.
    **Steps:** Shooting, Viewing
    **Characteristics:** Image quality, Essence

43. Use a wide shot to show the audience a wide view of the scene.
    **Rationale:** A wide shot enables you to establish the scene by showing the audience where the action is located.
    **Steps:** Shooting, Viewing
    **Characteristics:** Image quality, Essence

44. Prefer close to wide shots to create a video with more impact on the audience.
    **Rationale:** In contrast to wide shots, close shots add more emotion and drama to a video.
    **Steps:** Shooting, Viewing
    **Characteristics:** Image quality, Pleasure

45. Shoot scenes from different angles and heights to obtain a pleasant and interesting image.
    **Rationale:** Different angles and heights arouse the interest of the audience.

        **Steps:**        Shooting, Viewing
        **Characteristics:** Image quality, Pleasure

46. Avoid extreme angles to obtain a high-quality image.
    **Rationale:**      Extreme angles distort the image of your video.
    **Steps:**            Shooting
    **Characteristics:** Image quality

47. Do not shoot a subject in front of a black or strong colored (red, yellow, or bright-green) background to avoid a lower image quality.
    **Rationale:**      A black or strong colored background does not only distract the audience but such a background also modifies the apparent colors of the subject.
    **Steps:**            Shooting, Viewing
    **Characteristics:** Image quality, Clutter, Pleasure

48. When you record a group of persons with one camera, zoom out to include a new person in a wide shot and then zoom in on this person in a close shot to avoid continually panning across the group from one person to another.
    **Rationale:**      Any camera motion results in more compression of your video and can cause motion sickness by the audience.
    **Steps:**            Shooting, Viewing
    **Characteristics:** Image quality, Pleasure

49. Present a small object by placing it on a turntable (manual- or motor-driven) to avoid having to move the camera in an arc around the object.
    **Rationale:**      This motion is complex and therefore difficult to carry out smoothly.
    **Steps:**            Shooting
    **Characteristics:** Image quality

50. Carefully review the back- and foreground of a scene to ensure that there are no unplanned actions and objects included in a shot.
    **Rationale:**      Any unplanned action and object may distract the audience from the actual content that you want to convey.
    **Steps:**            Shooting, Viewing
    **Characteristics:** Clutter, Pleasure

51. Review the background of your scene for the subsequent factors to avoid inadvertently recording these factors.

    Factors to avoid in a shot:
    1. Reflections and contents of windows.
    2. Reflecting surfaces which may show the camera.
    3. Flashing signs, posters, directions signs, billboards, persons, etc. which may distract the audience.

    **Rationale:**      These factors can disrupt and distract the audience from the actual content of your video and thus must be excluded.
    **Steps:**            Shooting, Postproduction, Viewing
    **Characteristics:** Clutter, Pleasure

52. During the shooting, you have the subsequent options to achieve a suitable background for your scene.

    Options for the background:
    1. Rearrange the furniture.
    2. Replace the furniture with pieces from nearby rooms.
    3. Attach posters, notices, and signs to walls.

      **Rationale:**      These are quick, inexpensive, and simple options to customize the background of your scene to better suit your needs.
      **Steps:**      Shooting
**Characteristics:** Essence, Clutter

53. After a shot, check the sound quality and balance by listening to the recording with high-grade earphones or a loudspeaker to detect any unwanted background noises.
      **Rationale:**      Any unwanted background noise may disrupt and distract the audience from the actual content of your video.
      **Steps:**      Shooting, Viewing
**Characteristics:** Sound quality, Clutter, Pleasure

54. If something goes wrong during the recording of a scene, record the whole scene again to create one individual shot for one scene.
      **Rationale:**      Otherwise, the scene may not seem to be smooth and combining several shots for one scene increases your effort in the postproduction.
      **Steps:**      Shooting, Postproduction, Viewing
**Characteristics:** Image quality, Plot, Pleasure

55. Keep all shots, even the unsuccessful ones, to have a large collection of shots.
      **Rationale:**      Parts of, even the unsuccessful ones, may be used in the postproduction.
      **Steps:**      Shooting, Postproduction
**Characteristics:** Essence

56. Follow the subsequent steps of the non-linear editing process to create the final video.

    Steps of the non-linear editing process:
    1. Phase: Rough edit
        (a) Digitize footage on your computer to choose the best shots.
        (b) Trim and clean up each shot by deleting unwanted frames.
        (c) Place the shots on the timeline to assemble simply the structure of your video according to your planned story.
    2. Phase: Tight edit
        (a) Add effects and transitions to and between the shots.
        (b) Clean up and insert the necessary sound.
        (c) Before you create the final video, add titles to identify persons, places, and things supporting to the tell story and to give credits.
      **Rationale:**      This process is established in video production for highly structured, short-format videos due to its simplicity in making changes to the video by simply moving the shots around.
      **Steps:**      Postproduction
**Characteristics:** Plot, Clutter, Image quality, Sound quality

57. Keep the final video short (up to 5 minutes) to ensure that you only show the important details to the audience.
      **Rationale:**      You must focus the audience's attention on the important details which they need to understand.
      **Steps:**      Postproduction, Viewing
**Characteristics:** Video length, Essence, Pleasure

58. Do not create a rapid succession of unrelated shots or quick cuts between different viewpoints in the postproduction to avoid annoying, confusing, or boring your audience.
      **Rationale:**      The audience can hardly follow these presentation styles which in turn impede their understanding of the video.

    **Steps:** Postproduction, Viewing
**Characteristics:** Plot, Pleasure

59. Follow the subsequent rules for cutting in the postproduction to avoid irritating transitions in your final video.

    Rules for cutting:
    1. Plan to cut between shots as eye blinks when looking around.
    2. Do not cut between shots of extremely different sizes of the same subject, e.g., a close to a wide shot.
    3. Do not cut between shots that are similar or even match, e.g., two close shots of two different persons.
    4. Do not cut between two shots of the same size of the same subject, e.g., a close to a close shot.

    **Rationale:** These rules are based on experience and help you to ensure that your final video is pleasant to watch for your audience.
    **Steps:** Postproduction, Viewing
    **Characteristics:** Image quality, Pleasure

60. Include the subsequent graphical elements in your video to obtain a well structure.

    Graphical elements to structure your video:
    1. Opening title to announce the video.
    2. Subtitles to identify persons and locations.
    3. Credits to recognize the persons appearing in and contributing to the video.
    4. Ending titles to draw the video to its conclusion.

    **Rationale:** These graphical elements add clarity to your video and thus help the audience to follow the video and its content.
    **Steps:** Postproduction, Viewing
    **Characteristics:** Plot, Essence, Sense of responsibility, Pleasure

61. Consider the subsequent standards to design well-legible graphics.

    Standards for well-legible graphics:
    1. Place the graphic within the safe area, which is the 70 percent area around the center of the screen.
    2. Limit the number of fonts.
    3. Sans-serif bold fonts, e.g., Arial or Trade Gothic Bold, are best readable, especially on smaller displays.
    4. Avoid serif fonts since they create a flicker effect, especially on smaller displays.
    5. Letters smaller than one-tenth of the screen height are difficult to read.
    6. Black-edged letters are difficult to read.
    7. Do not use abbreviations to be unambiguous.
    8. Letters are usually much lighter than the background.
    9. Warm bright colors attract the most attention.

    **Rationale:** Well-legible graphics add clarity to the presentation of your video and thus help the audience to follow the video and its content.
    **Steps:** Postproduction, Viewing
    **Characteristics:** Image quality, Pleasure

62. When using materials (music, videos, images, texts, etc.) of third parties, verify that you comply with the respective regulations of copyright law to obtain a copy clearance for using the materials of third parties.

    **Rationale:** You are responsible and legally liable to comply with the legal regulations for copyright.
    **Steps:** Postproduction

**Characteristics:** Sense of responsibility

**63.** Label the final video by adding the subsequent metadata to have a fully labeled video.

Metadata:
1. Title.
2. Subtitle.
3. Department.
4. Producer/client.
5. Editor.
6. Video length.

**Rationale:** This information is important for future informational purposes, e.g., for distributing the video.
**Steps:** Postproduction
**Characteristics:** Sense of responsibility

# 3  Detailed Vision Video Production Guideline

Besides supporting the production and use of generic videos by software professionals, we are especially interested in supporting the production and use of vision videos due to our project-specific context. Based on our theoretical knowledge and practical experience gained in recent years in the production and use of vision videos in requirements engineering (cf. [1, 2, 4, 7, 8, 9, 10, 11, 12, 13, 14, 15, 16, 17, 18, 20, 22, 23, 24]), we derived the guideline for vision video production from the interdisciplinary guideline for video production. In particular, we adopted the recommendations of the interdisciplinary guideline whose usefulness we have experienced and thus can confirm for the production and use of vision videos. We revised the adopted recommendations by concretizing them in relation to vision videos. In addition, we also added new recommendations that we identified ourselves during the production and use of vision videos. We have to remark that we only included recommendations of the interdisciplinary guideline whose usefulness we could confirm, the other recommendations were omitted. This omission does not mean that the respective recommendations are not useful. We are currently not able to assess their relevance based on our knowledge and experience.

Out of the 63 recommendations of the interdisciplinary guideline for video production, we adopted and revised 55 recommendations that we could confirm based on our knowledge and experience. From these 55 recommendations, we derived 47 recommendations for vision videos by combining some of the recommendations for generic video into individual recommendations for vision videos. We also added five new recommendations that we identified ourselves during the production and use of vision videos. Thus, we present an experience-based guideline for vision video production that consists of 52 individual recommendations.

Below, we present each adopted or added recommendation in detail, i.e., the recommendation, its rationale, the affected steps of the video production and use process as well as the affected quality characteristics of vision videos. Furthermore, we indicate for each recommendation whether it is based on a recommendation of the interdisciplinary guideline or our own experience. In the case of more than one affected process step, we highlighted the step in which we mainly benefited from the respective recommendation. Below, we sorted the individual recommendations based on these highlighted steps along the video production and use process (italic highlighting). This sorting supports the practical applicability of the guideline since the recommendations are presented chronologically according to their required timely consideration in the video production and use process. In Appendix B – Vision Video Production Guideline, we present the experience-based guideline for vision video production that only contains the 52 recommendations sorted by the highlighted steps along the video production and use process.

**01.** Have preliminary meetings with your stakeholders to become sensitive to their concerns and different viewpoints.
  **Rationale:** These meetings help you to understand better your stakeholders by establishing trust between all parties involved.
  **Steps:** *Preproduction*
  **Characteristics:** Sense of responsibility, Clarity, Support
  **Based on:** Recommendation 01

02. Define the intended purpose and target audience of your vision video to clearly indicate your intention for what the vision video should be used for.

Intended purposes of vision videos:
1. Convey or obtain knowledge and/or new information (declarative knowledge) to share an integrated view of a future system and its use within a heterogeneous group of stakeholders for aligning their actions and views.
2. Convey or obtain knowledge and/or new information (declarative knowledge) to share an integrated view of the future system and its use with the development team that will implement the vision.
3. Convey or obtain particular experiences to share an integrated view of a future system and its use for validating this view and for eliciting new or diverging aspects. The vision video serves as a replacement of the future system and its use so that the viewers can experience the envisioned product.

**Rationale:** You must know the purpose and target audience for whom you want to produce a vision video to clearly indicate your intended use of the vision video for later viewing.
**Steps:** *Preproduction*, Viewing
**Characteristics:** Intention, Prior knowledge, Clarity, Support, Stability
**Based on:** Recommendation 02

03. Define the topics of a vision (addressed problem, key idea of the solution, and improvement of the problem by the solution) that you want to address in your vision video to clarify the content addressed in your vision video.
**Rationale:** These topics are crucial for a vision video since a vision video is a video that presents a vision or parts of it.
**Steps:** *Preproduction*, Viewing
**Characteristics:** Essence, Pleasure, Completeness
**Based on:** Recommendation 03

04. Keep your vision video simple by addressing a maximum of the three topics of a vision to reduce the amount of information in your vision video.
**Rationale:** The audience can understand a vision video with fewer topics easier than a vision video with too many topics.
**Steps:** *Preproduction*, Viewing
**Characteristics:** Essence, Pleasure, Focus, Completeness
**Based on:** Recommendation 04

05. Tell the content of your vision video by inventing a story with a beginning, middle, and end to create a clear structure of the contents of your vision video.
**Rationale:** This structure helps you to ensure that you introduce the prior knowledge (beginning) which the audience needs to know to understand the main content (middle) which has a defined conclusion (end).
**Steps:** *Preproduction*, Viewing
**Characteristics:** Plot, Prior knowledge, Completeness
**Based on:** Recommendation 07

06. If you are not sure how to invent the story of your vision video, you can use one of the subsequent storylines to tell the content of your vision video.

    1. Storyline:
        (a) Beginning: Address the audience emotionally by introducing the problem of your vision with its negative consequences.
        (b) Middle: Address the audience emotionally by introducing the key idea of the solution of your vision with its positive consequences.
        (c) End: Emphasize the envisioned improvement of the problem by the solution by concluding with its benefits.
    2. Storyline: (Requires that the audience knows the problem of the vision.)
        (a) Beginning: Introduce the key idea of the solution of your vision.
        (b) Middle: Emphasize the envisioned improvements of the solution.
        (c) End: Conclude with the benefits of your vision.

    **Rationale:** Both storylines have been applied in the production of vision videos. Based on experience, it can be confirmed that these two storylines work for vision videos.
    **Steps:** *Preproduction*, Viewing
    **Characteristics:** Plot, Prior knowledge, Pleasure, Focus, Completeness
    **Based on:** Experience

07. Use a storyboard, script, or narration to create an outline of the story of your vision video.

    A storyboard is a series of drawings that visualize the content of each scene used to plan the order of actions and events in the particular scene.

    Each element in the series consists of:
    1. An ID for each drawing.
    2. A hand-drawn sketch of a key image of the scene.
    3. A short textual description what happens in this part of the scene.
    4. If necessary, a textual description of the content of the audio track.

    **Rationale:** It is essential to plan the entire plot and main action of each scene in advance to organize the entire video production.
    **Steps:** *Preproduction*
    **Characteristics:** Plot, Essence
    **Based on:** Recommendation 08

08. In the case of recording a long action, plan to shoot or cut out only shots of strategic moments of the action to show its important moments in a condensed shot.
    **Rationale:** You must keep the duration of a shot short but at the same time show all important moments of a long action regardless of its length.
    **Steps:** *Preproduction*, Shooting, Viewing
    **Characteristics:** Plot, Video length
    **Based on:** Recommendation 13

09. Compose a shot by including a suitable back- and foreground for a scene to add additional information and a meaningful context to your vision video.

    Options for the back- and foreground of a scene:
    1. Use the real location.

2. Use a substitute for the location you need.
3. Build a set that resembles the real location.
4. Combine photos from the real location and sound effects with your shots to make the audience think that you were shooting in the real location.

**Rationale:** It can be difficult or impossible to shoot in the real location. In this case, you need to think about alternatives to illustrate the scene of your vision video properly. The back- and foreground must support the audience to understand the scene.
**Steps:** *Preproduction*, Shooting, Postproduction
**Characteristics:** Prior knowledge, Image quality, Essence
**Based on:** Recommendation 16, Recommendation 17

10. Create a list of all shots based on the storyboard by following the subsequent rules to plan the shooting order.

    Rules for planning the shooting:
    1. Create a list of one-liners that consist of the ID of the drawing of the storyboard and a short title of the respective shot.
    2. Sort the list by location.
    3. For each location, sort the shots again starting with shots that are easy to shoot and to understand for the actors.

    **Rationale:** You save time by shooting all shots of one location at once instead of shooting shots according to the running order of the plot.
    **Steps:** *Preproduction*, Shooting
    **Characteristics:** Plot, Sense of responsibility
    **Based on:** Recommendation 21

11. Ask the responsible authority of property you will use for your vision video to sign a consent form for shooting on the property to obtain the permission (signed consent form) to shoot on the property.
    **Rationale:** You are responsible and legally liable to comply with the legal regulations in the context of a video production.
    **Steps:** *Preproduction*, Shooting
    **Characteristics:** Sense of responsibility
    **Based on:** Recommendation 18

12. Ask each actor, even ad-hoc ones, to sign a consent form covering the subsequent topics to obtain the permission (signed consent form of each actor) to use and distribute the individual shots and the final vision video.

    Topics the consent form must cover:
    1. A statement of how all shots and the final vision video will be used and distributed.
    2. A statement that the actor has the right to withdraw at any time from the video production.
    3. A statement of whether the actor may request the deletion of the recordings at any time.
    4. A statement of how the actor can contact you (phone number or email).

    **Rationale:** You are responsible and legally liable to comply with the legal regulation of the general data protection regulation (GDPR).
    **Steps:** *Preproduction*, Shooting, Postproduction, Viewing
    **Characteristics:** Sense of responsibility
    **Based on:** Recommendation 19

13. Follow the subsequent steps to record a shot.

    Steps for recording:
    1. Review the details of the storyboard for the next scene.
    2. Call "Quiet, please!".
    3. Start recording.
    4. Call "Action!" to start the action.
    5. Call "Cut!" to announce the end of the action.
    6. Stop recording.
    7. Review the recording. If you are unsatisfied with the shot, repeat the recording of the entire shot.

    **Rationale:** These steps are an established process in video production for making a recording. It is important to record the entire shot again, otherwise, the scene may not appear smooth, and combining several shots for one scene increases your effort.
    **Steps:** *Shooting*, Postproduction, Viewing
    **Characteristics:** Essence, Clutter, Video length, Image quality, Sound quality, Plot, Pleasure
    **Based on:** Recommendation 22, Recommendation 54

14. Handle inexperienced actors by following the subsequent rules of conduct to calm the actors down.

    Rules of conduct for handling inexperienced actors:
    1. Making the actors feel welcome and that their contribution is important for the vision video.
    2. Give them instructions by telling them when, where, and how they should act in front of the camera.
    3. Make it clear to the actors that they do not have to worry that something goes wrong since the scene can be shot again.

    **Rationale:** Inexperienced actors are often nervous and afraid of doing something wrong that impedes shooting.
    **Steps:** *Shooting*
    **Characteristics:** Sense of responsibility
    **Based on:** Recommendation 23

15. Take the storyboard and sorted list of all shots with you to each scene during the shooting to know how to proceed.
    **Rationale:** The storyboard and sorted list of shots are your orientation for the shooting. Based on these two artifacts, you can plan your next shooting steps and check that each recorded shot corresponds to your expectations.
    **Steps:** *Shooting*
    **Characteristics:** Plot, Sense of responsibility, Completeness
    **Based on:** Experience

16. Before recording a shot, check the details of the storyboard and sorted list of all shots to ensure that you prepared everything.
    **Rationale:** There are often similar scenes. Therefore, you must check whether everything is prepared for a shot by having an explicit checklist. Otherwise, you can get easily confused.
    **Steps:** *Shooting*
    **Characteristics:** Plot, Sense of responsibility, Completeness
    **Based on:** Experience

17. Use the best camera available (smartphone, tablet, consumer HD camera, etc.) which is small and light and has a display, mounting option for an external microphone, a large storage capacity, and sockets for a connection with a computer to obtain high image quality and sound quality.
    **Rationale:** Today's video cameras offer an image quality and sound quality that is sufficient for the most purposes.
    **Steps:** *Shooting*
    **Characteristics:** Image quality, Sound quality
    **Based on:** Recommendation 24

18. Ensure that you have sufficient free memory capacity (SD cards) and at least two fully loaded batteries to avoid unnecessary interruptions during the shooting.
    **Rationale:** A full memory and an empty battery delay the shooting, thus increase your costs, and waste the time of all parties involved.
    **Steps:** *Shooting*
    **Characteristics:** Sense of responsibility
    **Based on:** Recommendation 35

19. In general, let the camera control auto-focus, white balance, and exposure controls to obtain a high image quality.
    **Rationale:** These camera settings are complex and have a strong impact on the image. Therefore, you should only adjust them manually if you have the necessary knowledge and experience.
    **Steps:** *Shooting*
    **Characteristics:** Image quality
    **Based on:** Recommendation 26

20. Use the standard recording speed to obtain a high image quality and sound quality.
    **Rationale:** The standard recording speed provides better image quality and sound quality.
    **Steps:** *Shooting*
    **Characteristics:** Image quality, Sound quality
    **Based on:** Recommendation 28

21. Use an external microphone, i.e., a shotgun microphone, to achieve high sound quality.
    **Rationale:** The built-in microphones rarely provide sufficient sound quality. A shotgun microphone is ideal for isolating a subject and eliminating nearby noises.
    **Steps:** *Shooting*
    **Characteristics:** Sound quality, Essence, Clutter
    **Based on:** Recommendation 25

22. For each shot, think about where you place the camera and microphone to create a vision video that enables the audience to experience the content of your vision video.
    **Rationale:** The more realistic the audience perceives the content of your vision video, the better they can understand the content of your vision video.
    **Steps:** *Shooting*, Viewing
    **Characteristics:** Image quality, Sound quality, Essence
    **Based on:** Recommendation 29

23. Have the light source, e.g., lamp or window, behind the camera to illuminate the subject.

      **Rationale:** The effect of light depends on the position of the camera. If you shoot against the light source, you leave the subject in deep shadow, making it difficult for the audience to recognize the subject.
      **Steps:** *Shooting*, Viewing
      **Characteristics:** Image quality, Essence
      **Based on:** Recommendation 31

24. Light the subject well by switching on the room lights or open the curtain to obtain a high image quality.
      **Rationale:** There is usually not enough light in a building to present a subject in such a way that the audience can recognize all details of the subject.
      **Steps:** *Shooting*, Viewing
      **Characteristics:** Image quality, Essence
      **Based on:** Recommendation 41

25. Place the microphone as close as possible to the subject by keeping the distance the same for all shots to get a higher sound quality.
      **Rationale:** The microphone is often too far away from the subject and strong reflections from nearby walls or loud background noises lower the quality of the sound.
      **Steps:** *Shooting*
      **Characteristics:** Sound quality, Clutter
      **Based on:** Recommendation 32

26. Compose a shot by keeping the important details of a scene within the safe area, which is the 70 percent area around the center of the screen, to ensure that the subject is properly framed and you do not accidentally cut off important details of the scene.
      **Rationale:** The audience might miss or not recognize the details that are outside of the safe area.
      **Steps:** Preproduction, *Shooting*, Viewing
      **Characteristics:** Image quality, Essence
      **Based on:** Recommendation 14

27. Compose a shot by following the subsequent standards of the rule of thirds, which divides the screen into thirds horizontally and vertically, to create a shot that looks dynamic.

    The rule of thirds:
    1. The subject should not be exactly in the middle of the screen.
    2. The subject should be on one of those lines and, ideally, on the intersection of two lines.
    3. If the subject moves towards something, position the subject behind the center of the screen according to the direction of the camera motion.
    4. If the subject moves away from something, position the subject over the center of the screen according to the direction of the camera motion.
    5. The faster the movement of the subject, the greater the offset from the center of the screen.

    **Rationale:** The rule of thirds helps you produce a nicely balanced image that attracts the interest of the audience.
    **Steps:** Preproduction, *Shooting*, Viewing
    **Characteristics:** Image quality, Essence, Pleasure
    **Based on:** Recommendation 15

28. Shoot scenes from different angles and heights to obtain a pleasant and interesting image.
    **Rationale:** Different angles and heights arouse the interest of the audience.
    **Steps:** *Shooting*, Viewing
    **Characteristics:** Image quality, Pleasure
    **Based on:** Recommendation 45

29. Avoid extreme angles to obtain a high-quality image.
    **Rationale:** Extreme angles distort the image of your vision video.
    **Steps:** *Shooting*
    **Characteristics:** Image quality
    **Based on:** Recommendation 46

30. Do not shoot a subject in front of a black or strong colored (red, yellow, or bright green) background to avoid a lower image quality.
    **Rationale:** A black or strong colored background does not only distract the audience but such a background also modifies the apparent colors of the subject.
    **Steps:** *Shooting*, Viewing
    **Characteristics:** Image quality, Clutter, Pleasure
    **Based on:** Recommendation 47

31. Review the back- and foreground of your scene for the subsequent factors to avoid inadvertently recording these factors.

    Factors to avoid in a shot:
    1. Reflections and contents of windows.
    2. Reflecting surfaces which may show the camera.
    3. Flashing signs, posters, directions signs, billboards, persons etc. which may distract the audience.

    **Rationale:** These factors can disrupt and distract the audience from the actual content of your vision video and thus must be excluded.
    **Steps:** *Shooting*, Postproduction, Viewing
    **Characteristics:** Clutter, Pleasure
    **Based on:** Recommendation 50, Recommendation 51

32. During the shooting, you have the subsequent options to achieve a suitable background for your scene.

    Options for the background:
    1. Rearrange the furniture.
    2. Replace the furniture with pieces from nearby rooms.
    3. Attach posters, notices, and signs to walls.

    **Rationale:** These are quick, inexpensive, and simple options to customize the background of your scene to better suit your needs.
    **Steps:** *Shooting*
    **Characteristics:** Essence, Clutter
    **Based on:** Recommendation 52

33. Stabilize the camera by using your body or a camera mount (monopod or tripod) to create a steady and carefully controlled image of your vision video.
    **Rationale:** If the image of your vision video is blurred, bounce around, or lean over to one side, it is a pain to watch for the audience.
    **Steps:** *Shooting*, Viewing
    **Characteristics:** Image quality, Pleasure

**Based on:** Recommendation 36

34. When using a tripod, turn the auto-focus off. Instead, position the camera and manually focus the image on a fixed point of interest.
    **Rationale:** If you use the auto-focus, any movement in the scene may cause the camera to adjust and re-adjust which in turn blurs the image for a moment.
    **Steps:** *Shooting*
    **Characteristics:** Image quality
    **Based on:** Recommendation 37

35. Sharpen the focus on the most important part of the scene and leave the rest defocused to highlight the important content of a scene.
    **Rationale:** Although this presentation style is not ideal, it enables you to show the audience exactly what you want them to see.
    **Steps:** *Shooting*, Viewing
    **Characteristics:** Image quality, Essence
    **Based on:** Recommendation 40

36. Use only as much camera motions, pans, zooms, and tilts as necessary by always ensuring that the motions are slow and smooth to create a shot with a low compression.
    **Rationale:** Any camera motion results in more compression of the image of a vision video and can cause motion sickness by the audience. Especially, zooming is an unnatural eye movement that can confuse the audience.
    **Steps:** *Shooting*, Viewing
    **Characteristics:** Image quality, Pleasure
    **Based on:** Recommendation 38, Recommendation 39

37. When you record a group of persons with one camera, zoom out to include a new person in a wide shot and then zoom in on this person in a close shot to avoid continually panning across the group from one person to another.
    **Rationale:** Any camera motion results in more compression of your video and can cause motion sickness by the audience.
    **Steps:** *Shooting*, Viewing
    **Characteristics:** Image quality, Pleasure
    **Based on:** Recommendation 48

38. Shoot every action from start to finish to ensure that the audience misses none of the action.
    **Rationale:** The audience gets an accurate impression of the action of a scene and its duration.
    **Steps:** *Shooting*, Viewing
    **Characteristics:** Plot, Essence, Video length, Pleasure
    **Based on:** Recommendation 33

39. Start the recording 5 seconds before the action starts and stop the recording 5 seconds after the action is completed to avoid too brief shots.
    **Rationale:** These additional buffers help you in the postproduction when you cut and combine the individual shots into the final vision video.
    **Steps:** *Shooting*, Postproduction
    **Characteristics:** Video length
    **Based on:** Recommendation 34

40. Hold each shot for at least 15 seconds and a maximum of 30 seconds to avoid to brief and too long shots.
    **Rationale:** These two times are the lower and upper boundary of the average storage period of the human short-term memory for capturing, processing, and understanding information. If you undercut respective exceed this duration, it is difficult for the audience to capture, process, understand, and remember the details of the shot.
    **Steps:** Preproduction, *Shooting*, Postproduction, Viewing
    **Characteristics:** Video length, Pleasure
    **Based on:** Recommendation 11, Recommendation 12
41. After recording a shot, check the image quality and sound quality by viewing the recording with high-grade earphones or a loudspeaker to detect any unwanted background or foreground actions and noises.
    **Rationale:** Any unwanted actions and noises may disrupt and distract the audience from the actual content of your vision video.
    **Steps:** *Shooting*, Viewing
    **Characteristics:** Image quality, Sound quality, Clutter, Pleasure
    **Based on:** Recommendation 50, Recommendation 53
42. Keep all shots, even the unsuccessful ones, to have a large collection of shots.
    **Rationale:** Parts of, even the unsuccessful shots, may be used in the postproduction.
    **Steps:** *Shooting*, Postproduction
    **Characteristics:** Essence
    **Based on:** Recommendation 55
43. Follow the subsequent steps of the non-linear editing process to create the final vision video.

    Steps of the non-linear editing process:
    1. Phase: Rough edit
        (a) Digitize footage on your computer to choose the best shots.
        (b) Trim and clean up each shot by deleting unwanted frames.
        (c) Place the shots on the timeline to assemble simply the structure of your vision video according to your planned story.
    2. Phase: Tight edit
        (a) Add effects and transitions to and between the shots.
        (b) Clean up and insert the necessary sound.
        (c) Before you create the final vision video, add titles to identify persons, places, and things supporting to the tell story and to give credits.

    **Rationale:** This process worked well for highly structured, short-format vision videos due to its simplicity in making changes to the vision video by simply moving the shots around.
    **Steps:** *Postproduction*
    **Characteristics:** Plot, Clutter, Image quality, Sound quality
    **Based on:** Recommendation 56
44. Ensure that the image- and soundtrack deal with the same topic to present consistent visual and auditive information.
    **Rationale:** The audience can be easily distracted and confused if the visual and auditive information of the image- and soundtrack do not match.
    **Steps:** Preproduction, Shooting, *Postproduction*, Viewing

**Characteristics:** Essence, Clutter, Pleasure
**Based on:** Recommendation 05, Recommendation 06

45. If the image- and soundtrack do not deal with the same topic, delete the soundtrack and maybe replace it to ensure that both tracks deal with the same topic.
    **Rationale:** It is easier to replace the soundtrack instead of the imagetrack. Therefore, you should always try to keep the image of a shot and replace the sound.
    **Steps:** *Postproduction*
    **Characteristics:** Essence, Clutter
    **Based on:** Experience

46. Do not create a rapid succession of unrelated shots or quick cuts between different viewpoints to avoid annoying, confusing, or boring your audience.
    **Rationale:** The audience can hardly follow these presentation styles which in turn impede their understanding of the vision video.
    **Steps:** *Postproduction*, Viewing
    **Characteristics:** Plot, Pleasure
    **Based on:** Recommendation 58

47. Follow the subsequent rules for cutting in the postproduction to avoid irritating transitions in your final vision video.

    Rules for cutting:
    1. Plan to cut between shots as eye blinks when looking around.
    2. Do not cut between shots of extremely different sizes of the same subject, e.g., a close to a wide shot.
    3. Do not cut between shots that are similar or even match, e.g., two close shots of two different persons.
    4. Do not cut between two shots of the same size of the same subject, e.g., a close to a close shot.

    **Rationale:** These rules are based on experience and help you to ensure that your final vision video is pleasant to watch for your audience.
    **Steps:** *Postproduction*, Viewing
    **Characteristics:** Image quality, Pleasure
    **Based on:** Recommendation 59

48. Include the subsequent graphical elements in your vision video to obtain a well structure.

    Graphical elements to structure your vision video:
    1. Opening title to announce the vision video.
    2. Subtitles to identify persons and locations.
    3. Credits to recognize the persons appearing in and contributing to the vision video.
    4. Ending titles to draw the vision video to its conclusion.

    **Rationale:** These graphical elements add clarity to your vision video and thus help the audience to follow the vision video and its content.
    **Steps:** *Postproduction*, Viewing
    **Characteristics:** Plot, Essence, Sense of responsibility, Pleasure
    **Based on:** Recommendation 60

49. When using materials (music, videos, images, texts, etc.) of third parties, verify that you comply with the respective regulations of copyright law to obtain a copy clearance for using the materials of third parties.

    **Rationale:**    You are responsible and legally liable to comply with the legal regulations for copyright.
    **Steps:**    *Postproduction*
    **Characteristics:**    Sense of responsibility
    **Based on:**    Recommendation 62

50. Consider the subsequent standards to design well-legible graphics.

    Standards for well-legible graphics:
    1. Place the graphic within the safe area, which is the 70 percent area around the center of the screen.
    2. Limit the number of fonts.
    3. Sans-serif bold fonts, e.g., Arial or Trade Gothic Bold, are best readable, especially on smaller displays.
    4. Avoid serif fonts since they create a flicker effect, especially on smaller displays.
    5. Letters smaller than one-tenth of the screen height are difficult to read.
    6. Black-edged letters are difficult to read.
    7. Do not use abbreviations to be unambiguous.
    8. Letters are usually much lighter than the background.
    9. Warm bright colors attract the most attention.

    **Rationale:**    Well-legible graphics add clarity to the presentation of your vision video and thus help the audience to follow the vision video and its content.
    **Steps:**    *Postproduction*, Viewing
    **Characteristics:**    Image quality, Pleasure
    **Based on:**    Recommendation 61

51. Keep the final vision video short (up to 5 minutes) to ensure that you only show the important details to the audience.
    **Rationale:**    You must focus the audience's attention on the important details which they need to understand. The duration of a shot is crucial. If a shot is too long, the audience loses interest since they cannot capture, process, and understand the information presented. If a shot is too brief, the audience captures, but cannot process and understand, the information presented.
    **Steps:**    Preproduction, Shooting, *Postproduction*, Viewing
    **Characteristics:**    Video length, Essence, Pleasure, Focus
    **Based on:**    Recommendation 10, Recommendation 57

52. Before you produce the final vision video, ask a second person to review your vision video for comprehensibility as well as visual and auditory problems.
    **Rationale:**    You must exclude blurred images, poor sound, and tacit assumptions which often tend to be unnoticed by you as the video producer. Ask someone to review the vision video to see if this person can follow the logic and speed of the story of your vision video.
    **Steps:**    *Postproduction*, Viewing
    **Characteristics:**    Image quality, Sound quality, Plot, Essence, Pleasure, Clarity, Completeness
    **Based on:**    Experience

# Appendix A – Video Production Guideline

1. Have preliminary meetings with your stakeholders to become sensitive to their concerns and different viewpoints.
2. Define the intended purpose and target audience of your video to plan the content of your video.
3. Create a list of topics that you want to address in your video to define the final number of topics addressed in your video.
4. Keep your video simple by addressing only a few topics to reduce the amount of information in your video.
5. Deal with one topic at a time to avoid cuts between different topics, flashbacks, and flashforwards.
6. Ensure that the image- and soundtrack deal with the same topic to present consistent visual and auditive information.
7. Tell the content of your video as a story that has a beginning, middle, and end to create a clear structure of the contents of your video.
8. Use a storyboard, script, or narration to create an outline of the story of your video.
9. Plan the plot of your video by designing the scenes in such a way that they can be trimmed or omitted as needed to shorten the final video if necessary.
10. Think about how long a shot should last to define the final duration of the shot.
11. Hold a shot for at least 15 seconds to enable the audience to understand the information presented.
12. Hold a shot for a maximum of 30 seconds to avoid too long shots that the audience cannot understand.
13. Plan to shoot shots of strategic moments of a long action to show the important moments of the long action in a condensed shot.
14. Compose a shot by keeping the important details of a scene within the safe area, which is the 70 percent area around the center of the screen, to ensure that the subject is properly framed and you do not accidentally cut off important details of the scene.
15. Compose a shot by following the subsequent standards of the rule of thirds, which divides the screen into thirds horizontally and vertically, to create a shot that looks dynamic.
16. Compose a shot by including a suitable back- and foreground for a scene to add additional information and a meaningful context to your video.
17. You have the subsequent options to obtain a suitable background for your scene.

    Options for the background of a scene:
    1. Use the real location.
    2. Use a substitute for the location you need.
    3. Build a set that resembles the real location.
    4. Combine photos from the real location and sound effects with your shots to make the audience think that you were shooting in the real location.

18. Ask the owner or responsible authority of private or public property to sign a consent form for shooting on the property to obtain the permission (signed consent form) to shoot on the property.

19. Ask each actor to sign a consent form covering the subsequent topics to obtain the permission (signed consent form of each actor) to use and distribute the individual shots and the final video.

    Topics the consent form must cover:
    1. A statement of how all shots and the final video will be used and distributed.
    2. A statement that the actor has the right to withdraw at any time from the video production.
    3. A statement of whether the actor may request the deletion of the recordings at any time.
    4. A statement of how the actor can contact you (phone number or email).

20. Give each actor a suitable amount of time for reading the consent form as well as asking questions and clarifying issues and concerns to achieve that the actor is completely informed about the video production and use.

21. Plan to shoot all of the action at one location before you go to the next location to simplify the shooting.

22. Follow the subsequent steps to record a shot.

    Steps for recording:
    1. Review the details of the storyboard for the next scene.
    2. Call "Quiet, please!".
    3. Start recording.
    4. Call "Action!" to start the action.
    5. Call "Cut!" to announce the end of the action.
    6. Stop recording.
    7. Review the recording. If you are unsatisfied with the shot, repeat the recording.

23. Handle inexperienced actors by following the subsequent rules of conduct to calm the actors down.

    Rules of conduct for handling inexperienced actors:
    1. Making the actors feel welcome and that their contribution is important for the video.
    2. Give them instructions by telling them when, where, and how they should act in front of the camera.
    3. Make it clear to the actors that they do not have to worry that something goes wrong since the scene can be shot again.

24. Use the best camera available (smartphone, tablet, consumer HD camera, etc.) which is small and light and has a display, mounting option for an external microphone, a large storage capacity, and sockets for a connection with a computer to obtain high image and sound quality.

25. Use an external microphone, i.e., a shotgun microphone, to achieve high sound quality.

26. In general, let the camera control auto-focus, white balance, and exposure controls to obtain a high image quality.

27. Use the 720p (1280 x 720) or better 1080p (1920 x 1080) HD format which utilizes a 16:9 aspect ratio to obtain a high-quality image.

28. Use the standard recording speed to obtain high-quality image and sound.

29. Consider where you place the camera and microphone to create a video that enables the audience to experience the content of your video.

30. When using the camera, keep in mind that you restrict the perspective of the image the audience can see in the video to avoid frustrating the audience.

31. Have the light source, e.g., lamp or window, behind the camera to illuminate the subject.

32. Place the microphone as close as possible to the subject by keeping the distance the same for all shots to get a higher sound quality.

33. For brief video production, shoot every action from start to finish to ensure that the audience misses none of the action.

34. Start the recording 5 – 10 seconds before the action starts and stop the recording 5 – 10 seconds after the action is completed to avoid too brief shots.

35. Watch the elapsed time on the camera to know how much memory and battery are left.

36. Stabilize the camera by using your body or a camera mount (monopod or tripod) to create a steady and carefully controlled image of your video.

37. When using a tripod, turn the auto-focus off. Instead, position the camera and manually focus the image on a fixed point of interest.

38. Use only as much camera motions, pans, zooms, and tilts as necessary by always ensuring that the motions are slow and smooth to create a shot with a low compression.

39. If possible, do not zoom in or out to obtain a high image quality of your video.

40. Sharpen the focus on the most important part of the scene and leave the rest defocused to highlight the important content of a scene.

41. Light the subject well by switching on the room lights or open the curtain to obtain a high image quality.

42. Use a close shot to show the audience the action of a scene.

43. Use a wide shot to show the audience a wide view of the scene.

44. Prefer close to wide shots to create a video with more impact on the audience.

45. Shoot scenes from different angles and heights to obtain a pleasant and interesting image.

46. Avoid extreme angles to obtain a high-quality image.

47. Do not shoot a subject in front of a black or strong colored (red, yellow, or bright-green) background to avoid a lower image quality.

48. When you record a group of persons with one camera, zoom out to include a new person in a wide shot and then zoom in on this person in a close shot to avoid continually panning across the group from one person to another.

49. Present a small object by placing it on a turntable (manual- or motor-driven) to avoid having to move the camera in an arc around the object.

50. Carefully review the back- and foreground of a scene to ensure that there are no unplanned actions and objects included in a shot.

51. Review the background of your scene for the subsequent factors to avoid inadvertently recording these factors.

    Factors to avoid in a shot:
    1. Reflections and contents of windows.
    2. Reflecting surfaces which may show the camera.
    3. Flashing signs, posters, directions signs, billboards, persons, etc. which may distract the audience.

52. During the shooting, you have the subsequent options to achieve a suitable background for your scene.

    Options for the background:
    1. Rearrange the furniture.
    2. Replace the furniture with pieces from nearby rooms.
    3. Attach posters, notices, and signs to walls.

53. After a shot, check the sound quality and balance by listening to the recording with high-grade earphones or a loudspeaker to detect any unwanted background noises.

54. If something goes wrong during the recording of a scene, record the whole scene again to create one individual shot for one scene.

55. Keep all shots, even the unsuccessful ones, to have a large collection of shots.

56. Follow the subsequent steps of the non-linear editing process to create the final video.

    Steps of the non-linear editing process:
    1. Phase: Rough edit
        (a) Digitize footage on your computer to choose the best shots.
        (b) Trim and clean up each shot by deleting unwanted frames.
        (c) Place the shots on the timeline to assemble simply the structure of your video according to your planned story.
    2. Phase: Tight edit
        (a) Add effects and transitions to and between the shots.
        (b) Clean up and insert the necessary sound.
        (c) Before you create the final video, add titles to identify persons, places, and things supporting to the tell story and to give credits.

57. Keep the final video short (up to 5 minutes) to ensure that you only show the important details to the audience.

58. Do not create a rapid succession of unrelated shots or quick cuts between different viewpoints in the postproduction to avoid annoying, confusing, or boring your audience.

59. Follow the subsequent rules for cutting in the postproduction to avoid irritating transitions in your final video.
    Rules for cutting:
    1. Plan to cut between shots as eye blinks when looking around.
    2. Do not cut between shots of extremely different sizes of the same subject, e.g., a close to a wide shot.
    3. Do not cut between shots that are similar or even match, e.g., two close shots of two different persons.
    4. Do not cut between two shots of the same size of the same subject, e.g., a close to a close shot.

60. Include the subsequent graphical elements in your video to obtain a well structure.
    Graphical elements to structure your video:
    1. Opening title to announce the video.
    2. Subtitles to identify persons and locations.
    3. Credits to recognize the persons appearing in and contributing to the video.
    4. Ending titles to draw the video to its conclusion.

61. Consider the subsequent standards to design well-legible graphics.
    Standards for well-legible graphics:
    1. Place the graphic within the safe area, which is the 70 percent area around the center of the screen.
    2. Limit the number of fonts.
    3. Sans-serif bold fonts, e.g., Arial or Trade Gothic Bold, are best readable, especially on smaller displays.
    4. Avoid serif fonts since they create a flicker effect, especially on smaller displays.
    5. Letters smaller than one-tenth of the screen height are difficult to read.
    6. Black-edged letters are difficult to read.
    7. Do not use abbreviations to be unambiguous.
    8. Letters are usually much lighter than the background.
    9. Warm bright colors attract the most attention.

62. When using materials (music, videos, images, texts, etc.) of third parties, verify that you comply with the respective regulations of copyright law to obtain a copy clearance for using the materials of third parties.

63. Label the final video by adding the subsequent metadata to have a fully labeled video.
    Metadata:
    (a) Title.
    (b) Subtitle.
    (c) Department.
    (d) Producer/client.
    (e) Editor.
    (f) Video length.

# Appendix B – Vision Video Production Guideline

1. Have preliminary meetings with your stakeholders to become sensitive to their concerns and different viewpoints.

2. Define the intended purpose and target audience of your vision video to clearly indicate your intention for what the vision video should be used for.

    Intended purposes of vision videos:
    1. Convey or obtain knowledge and/or new information (declarative knowledge) to share an integrated view of a future system and its use within a heterogeneous group of stakeholders for aligning their actions and views.
    2. Convey or obtain knowledge and/or new information (declarative knowledge) to share an integrated view of the future system and its use with the development team that will implement the vision.
    3. Convey or obtain particular experiences to share an integrated view of a future system and its use for validating this view and for eliciting new or diverging aspects. The vision video serves as a replacement of the future system and its use so that the viewers can experience the envisioned product.

3. Define the topics of a vision (addressed problem, key idea of the solution, and improvement of the problem by the solution) that you want to address in your vision video to clarify the content addressed in your vision video.

4. Keep your vision video simple by addressing a maximum of the three topics of a vision to reduce the amount of information in your vision video.

5. Tell the content of your vision video by inventing a story with a beginning, middle, and end to create a clear structure of the contents of your vision video.

6. If you are not sure how to invent the story of your vision video, you can use one of the subsequent storylines to tell the content of your vision video.
    1. Storyline:
        (a) Beginning: Address the audience emotionally by introducing the problem of your vision with its negative consequences.
        (b) Middle: Address the audience emotionally by introducing the key idea of the solution of your vision with its positive consequences.
        (c) End: Emphasize the envisioned improvement of the problem by the solution by concluding with its benefits.
    2. Storyline: (Requires that the audience knows the problem of the vision.)
        (a) Beginning: Introduce the key idea of the solution of your vision.
        (b) Middle: Emphasize the envisioned improvements of the solution.
        (c) End: Conclude with the benefits of your vision.

7. Use a storyboard, script, or narration to create an outline of the story of your vision video.

    A storyboard is a series of drawings that visualize the content of each scene used to plan the order of actions and events in the particular scene.

    Each element in the series consists of:

1. An ID for each drawing.
2. A hand-drawn sketch of a key image of the scene.
3. A short textual description what happens in this part of the scene.
4. If necessary, a textual description of the content of the audio track.

8. In the case of recording a long action, plan to shoot or cut out only shots of strategic moments of the action to show its important moments in a condensed shot.
9. Compose a shot by including a suitable back- and foreground for a scene to add additional information and a meaningful context to your vision video.

    Options for the back- and foreground of a scene:
    1. Use the real location.
    2. Use a substitute for the location you need.
    3. Build a set that resembles the real location.
    4. Combine photos from the real location and sound effects with your shots to make the audience think that you were shooting in the real location.

10. Create a list of all shots based on the storyboard by following the subsequent rules to plan the shooting order.

    Rules for planning the shooting:
    1. Create a list of one-liners that consist of the ID of the drawing of the storyboard and a short title of the respective shot.
    2. Sort the list by location.
    3. For each location, sort the shots again starting with shots that are easy to shoot and to understand for the actors.

11. Ask the responsible authority of property you will use for your vision video to sign a consent form for shooting on the property to obtain the permission (signed consent form) to shoot on the property.
12. Ask each actor, even ad-hoc ones, to sign a consent form covering the subsequent topics to obtain the permission (signed consent form of each actor) to use and distribute the individual shots and the final vision video.

    Topics the consent form must cover:
    1. A statement of how all shots and the final vision video will be used and distributed.
    2. A statement that the actor has the right to withdraw at any time from the video production.
    3. A statement of whether the actor may request the deletion of the recordings at any time.
    4. A statement of how the actor can contact you (phone number or email).

13. Follow the subsequent steps to record a shot.

    Steps for recording:
    1. Review the details of the storyboard for the next scene.
    2. Call "Quiet, please!".
    3. Start recording.

4. Call "Action!" to start the action.
   5. Call "Cut!" to announce the end of the action.
   6. Stop recording.
   7. Review the recording. If you are unsatisfied with the shot, repeat the recording of the entire shot.
14. Handle inexperienced actors by following the subsequent rules of conduct to calm the actors down.

    Rules of conduct for handling inexperienced actors:
    1. Making the actors feel welcome and that their contribution is important for the vision video.
    2. Give them instructions by telling them when, where, and how they should act in front of the camera.
    3. Make it clear to the actors that they do not have to worry that something goes wrong since the scene can be shot again.
15. Take the storyboard and sorted list of all shots with you to each scene during the shooting to know how to proceed.
16. Before recording a shot, check the details of the storyboard and sorted list of all shots to ensure that you prepared everything.
17. Use the best camera available (smartphone, tablet, consumer HD camera, etc.) which is small and light and has a display, mounting option for an external microphone, a large storage capacity, and sockets for a connection with a computer to obtain high image quality and sound quality.
18. Ensure that you have sufficient free memory capacity (SD cards) and at least two fully loaded batteries to avoid unnecessary interruptions during the shooting.
19. In general, let the camera control auto-focus, white balance, and exposure controls to obtain a high image quality.
20. Use the standard recording speed to obtain a high image quality and sound quality.
21. Use an external microphone, i.e., a shotgun microphone, to achieve high sound quality.
22. For each shot, think about where you place the camera and microphone to create a vision video that enables the audience to experience the content of your vision video.
23. Have the light source, e.g., lamp or window, behind the camera to illuminate the subject.
24. Light the subject well by switching on the room lights or open the curtain to obtain a high image quality.
25. Place the microphone as close as possible to the subject by keeping the distance the same for all shots to get a higher sound quality.
26. Compose a shot by keeping the important details of a scene within the safe area, which is the 70 percent area around the center of the screen, to ensure that the subject is properly framed and you do not accidentally cut off important details of the scene.
27. Compose a shot by following the subsequent standards of the rule of thirds, which divides the screen into thirds horizontally and vertically, to create a shot that looks dynamic.

The rule of thirds:
1. The subject should not be exactly in the middle of the screen.
2. The subject should be on one of those lines and, ideally, on the intersection of two lines.
3. If the subject moves towards something, position the subject behind the center of the screen according to the direction of the camera motion.
4. If the subject moves away from something, position the subject over the center of the screen according to the direction of the camera motion.
5. The faster the movement of the subject, the greater the offset from the center of the screen.

28. Shoot scenes from different angles and heights to obtain a pleasant and interesting image.
29. Avoid extreme angles to obtain a high-quality image.
30. Do not shoot a subject in front of a black or strong colored (red, yellow, or bright green) background to avoid a lower image quality.
31. Review the back- and foreground of your scene for the subsequent factors to avoid inadvertently recording these factors.

    Factors to avoid in a shot:
    1. Reflections and contents of windows.
    2. Reflecting surfaces which may show the camera.
    3. Flashing signs, posters, directions signs, billboards, persons etc. which may distract the audience

32. During the shooting, you have the subsequent options to achieve a suitable background for your scene.

    Options for the background:
    1. Rearrange the furniture.
    2. Replace the furniture with pieces from nearby rooms.
    3. Attach posters, notices, and signs to walls.

33. Stabilize the camera by using your body or a camera mount (monopod or tripod) to create a steady and carefully controlled image of your vision video.
34. When using a tripod, turn the auto-focus off. Instead, position the camera and manually focus the image on a fixed point of interest.
35. Sharpen the focus on the most important part of the scene and leave the rest defocused to highlight the important content of a scene.
36. Use only as much camera motions, pans, zooms, and tilts as necessary by always ensuring that the motions are slow and smooth to create a shot with a low compression.
37. When you record a group of persons with one camera, zoom out to include a new person in a wide shot and then zoom in on this person in a close shot to avoid continually panning across the group from one person to another.

38. Shoot every action from start to finish to ensure that the audience misses none of the action.
39. Start the recording 5 seconds before the action starts and stop the recording 5 seconds after the action is completed to avoid too brief shots.
40. Hold each shot for at least 15 seconds and a maximum of 30 seconds to avoid to brief and too long shots.
41. After recording a shot, check the image quality and sound quality by viewing the recording with high-grade earphones or a loudspeaker to detect any unwanted background or foreground actions and noises.
42. Keep all shots, even the unsuccessful ones, to have a large collection of shots.
43. Follow the subsequent steps of the non-linear editing process to create the final vision video.

    Steps of the non-linear editing process:
    1. Phase: Rough edit
        (a) Digitize footage on your computer to choose the best shots.
        (b) Trim and clean up each shot by deleting unwanted frames.
        (c) Place the shots on the timeline to assemble simply the structure of your vision video according to your planned story.
    2. Phase: Tight edit
        (a) Add effects and transitions to and between the shots.
        (b) Clean up and insert the necessary sound.
        (c) Before you create the final vision video, add titles to identify persons, places, and things supporting to the tell story and to give credits.

44. Ensure that the image- and soundtrack deal with the same topic to present consistent visual and auditive information.
45. If the image- and soundtrack do not deal with the same topic, delete the soundtrack and maybe replace it to ensure that both tracks deal with the same topic.
46. Do not create a rapid succession of unrelated shots or quick cuts between different viewpoints to avoid annoying, confusing, or boring your audience.
47. Follow the subsequent rules for cutting in the postproduction to avoid irritating transitions in your final vision video.

    Rules for cutting:
    1. Plan to cut between shots as eye blinks when looking around.
    2. Do not cut between shots of extremely different sizes of the same subject, e.g., a close to a wide shot.
    3. Do not cut between shots that are similar or even match, e.g., two close shots of two different persons.
    4. Do not cut between two shots of the same size of the same subject, e.g., a close to a close shot.

48. Include the subsequent graphical elements in your vision video to obtain a well structure.

    Graphical elements to structure your vision video:
    1. Opening title to announce the vision video.
    2. Subtitles to identify persons and locations.
    3. Credits to recognize the persons appearing in and contributing to the vision video.
    4. Ending titles to draw the vision video to its conclusion.

49. When using materials (music, videos, images, texts, etc.) of third parties, verify that you comply with the respective regulations of copyright law to obtain a copy clearance for using the materials of third parties.

50. Consider the subsequent standards to design well-legible graphics.

    Standards for well-legible graphics:
    1. Place the graphic within the safe area, which is the 70 percent area around the center of the screen.
    2. Limit the number of fonts.
    3. Sans-serif bold fonts, e.g., Arial or Trade Gothic Bold, are best readable, especially on smaller displays.
    4. Avoid serif fonts since they create a flicker effect, especially on smaller displays.
    5. Letters smaller than one-tenth of the screen height are difficult to read.
    6. Black-edged letters are difficult to read.
    7. Do not use abbreviations to be unambiguous.
    8. Letters are usually much lighter than the background.
    9. Warm bright colors attract the most attention.

51. Keep the final vision video short (up to 5 minutes) to ensure that you only show the important details to the audience.

52. Before you produce the final vision video, ask a second person to review your vision video for comprehensibility as well as visual and auditory problems.